# 389.3-Tb/s 1017-km C-band Transmission over Field-Installed 12-Coupled-Core Fiber Cable with >12-Tb/s Spatial MIMO Channels

Akira Kawai[(1)], Kohki Shibahara[(1)], Megumi Hoshi[(1)], Masanori Nakamura[(1)], Takayuki Kobayashi[(1)], Ryota Imada[(2)], Takayoshi Mori[(2)], Taiji Sakamoto[(2)], Yusuke Yamada[(2)], Kazuhide Nakajima[(2)], Munehiko Nagatani[(1,3)], Hitoshi Wakita[(3)], Yuta Shiratori[(3)], Hiroshi Yamazaki[(1,3)], Hiroyuki Takahashi[(1,3)], Soichi Endo[(4)], Takemi Hasegawa[(4)], Ryo Nagase[(5)], and Yutaka Miyamoto[(1)]

[(1)] NTT Network Innovation Laboratories, NTT Corporation, Yokosuka, Japan, akira.kawai@ntt.com
[(2)] NTT Access Service Systems Laboratories, NTT Corporation, Tsukuba, Japan
[(3)] NTT Device Technology Laboratories, NTT Corporation, Atsugi, Japan
[(4)] Optical Communications Laboratory, Sumitomo Electric Industries, Yokohama, Japan
[(5)] Faculty of Engineering, Chiba Institute of Technology, Narashino, Japan

**Abstract** *We demonstrate 4.65-THz WDM/SDM transmission of 140-Gbaud PS-QAM signals over field-installed 12-coupled-core fiber cable with standard cladding diameter, achieving a record 0.455 Pb/s coupled-core capacity in a field environment. We also demonstrate 0.389 Pb/s over-1000-km transmission of spatial MIMO channels with >12 Tb/s/wavelength net bitrate. ©2024 The Authors*

**Introduction**

Space-division multiplexing (SDM) systems leverage multiple spatial degrees of freedom of optical fibers, offering the potential for ultra-high transmission capacity per fiber [1]. Coupled-core multicore fibers (CC-MCF) [2,3], notably, enable the multiplexing of more than ten channels within the standard cladding diameter, with low fiber nonlinearity and spatial-mode dispersion (SMD).

To date, laboratory experiments have reported coupled-core transmission capacities of up to 1.7 Pb/s (19 cores, C+L bands) [4], while field demonstrations have achieved about 21 Tb/s (4 cores, C band) [5,6]. Toward deployment of CC-MCF in 1000-km-class high-capacity terrestrial systems, two challenges have been remained. The first is to validate the feasibility of high-capacity transmission in field-installed cable environments with many connection points. The second is to investigate fully spatially coupled multiple-input/multiple-output (MIMO) transmission using over 100-Gbaud broadband signals.

In this work, we report for the first time the results of transmission experiments over a field-installed 12-coupled-core fiber (12CCF) cable-based transmission line including multiple multicore splicing and connector points. Using a 53.5-km span transmission line and a total of 4.65 THz high-symbol-rate 140-Gbaud (150-GHz grid) WDM signals, we demonstrated high-capacity field transmission of net 455.4 Tb/s over a single-span 53.5-km transmission line, and net 389.3 Tb/s over a 19-span, 1017-km transmission line using only 31 wavelength channels. These results were achieved using a high-speed transmitter utilizing broadband InP-double heterojunction bipolar transistor (DHBT) bandwidth (BW) doublers, 12CCF with low SMD (<6 ps/km$^{1/2}$), and a frequency-domain complex 96×24 MIMO adaptive equalizer, achieving net bitrates of >12 Tb/s/wavelength with spatially coupled 24-channel MIMO processing. Figure 1 shows a comparison with recent C-band SDM transmission experiments using standard cladding diameter fibers [6,7,8,9,10]. Notably, this work represents the highest transmission capacity in a terrestrial field environment.

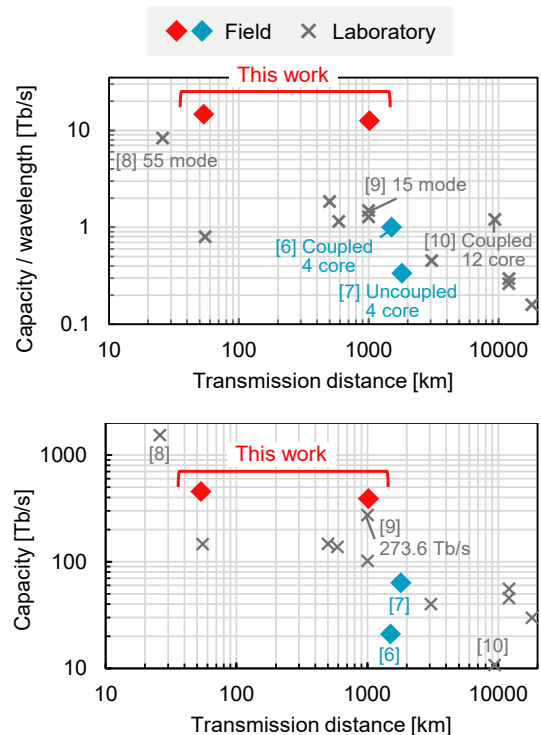

**Fig. 1:** Comparison of this work to recent C-band transmission experiments with standard cladding diameter SDM fibers

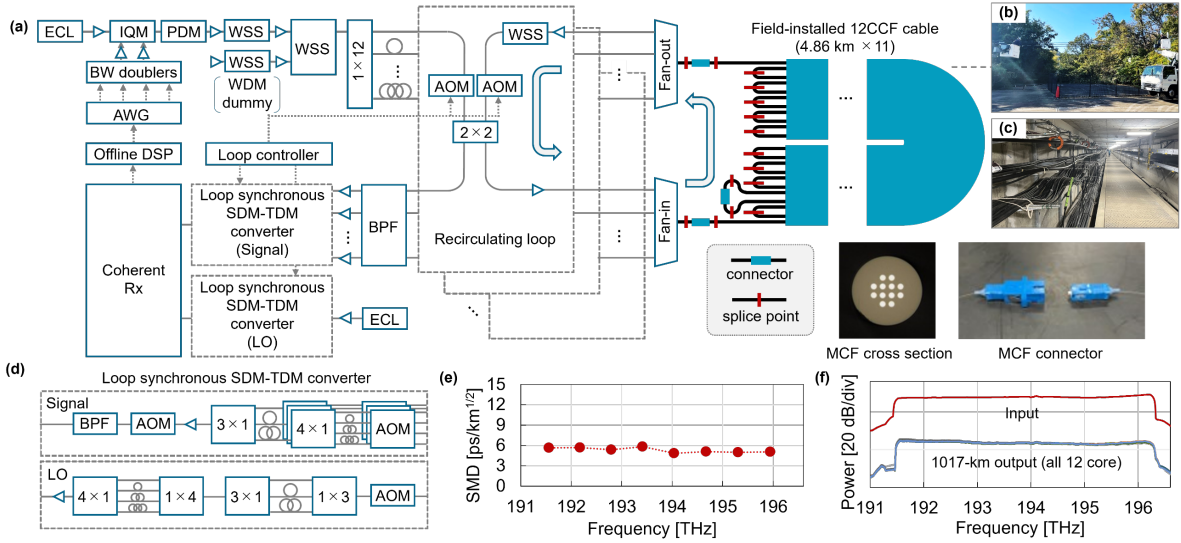

**Fig. 2 :** (a) Experimental setup. (b) Aerial installation of the fiber cable. (c) The fiber cable installed in an underground cable tunnel. (d) SDM-TDM converter. (e) SMD of the installed fiber. (h) WDM spectrum.

**Experimental setup**

The schematic illustration of the experiment is shown in Fig. 2. The cable is set up in a 200-fiber configuration utilizing technology similar to that described in a previous study [11], which enables a flat and low SMD over the entire C-band in a ~5-km-long cable with a cut-off wavelength lower than 1530 nm and G.652.D-compliant bending loss [12]. An overview of the experimental setup is provided in Fig. 2(a), and parameters of the transmission line are summarized in Tab. 1. The total length of the cable is 4.86 km, including a ~0.2 km aerial section as shown in Fig. 2(b) and a total of ~4.1 km underground tunnel sections as shown in Fig. 2(c). The SMD of the fiber was minimized to an average of 5.3 ps/km$^{1/2}$ across the entire C-band as shown in Fig. 2(e). A transmission line of 53.5 km was constructed by directly interconnecting of 11 fibers in the cable with MCF splicing and MCF connectors. The MCF connectors conformed to the SC standard [13] (see photograph in Fig. 2(a)).

The 140-Gbaud coherent signals were generated using a 4-ch. 32 GHz CMOS DAC-based arbitrary waveform generator (AWG) and two 89-GHz BW doublers with digital inverse multiplexing-based pre-distortion [14]. The signal format was a truncated probabilistically shaped 36-ary quadrature amplitude modulation (PS-36QAM), derived from PS-64QAM, with a 2D entropy of 4.688 bits. Continuous waves from a 10-kHz external cavity laser (ECL) were modulated by a 22-GHz In-phase/quadrature modulator (IQM) and polarization division multiplexing (PDM) was then emulated using polarization beam splitters and a delay line. This was followed by frequency response compensation of the transmitter using a wavelength-selective switch (WSS)-based optical equalization. The signal under test (SUT) was then combined with a WDM dummy based on amplified spontaneous emission noise from an erbium-doped fiber amplifier (EDFA) via another WSS. Core multiplexing was emulated by a 1×12 splitter and delay lines, and each spatial tributary entered 12 parallel recirculating loops. The loops were interfaced with a fan-in and fan-out connected to the transmission line. Each loop included EDFAs placed before and after the transmission line, along with WSSs for the gain equalization. The input power per core was 20.0 dBm and the total fiber input power was 30.8 dBm. Optical spectra before and after transmission are shown in Fig. 2(f). Polarization scramblers were not used in this work as they do not affect the transmission quality in strongly coupled coupled-core transmissions with a recirculating loop configuration [15]. The SUT was isolated by band-pass filters (BPF) before signal reception.

The signal reception system adopted an SDM-TDM (time-division multiplexed) configuration [16], in which 12-core multiplexed signals were received by a single high-speed coherent receiver (Rx). The SUT was injected into an SDM-TDM converter, which mainly consisted of acousto-optic modulators (AOM) and delay lines as shown in Fig. 2(d) and was synchronized with the recirculation loops. Each spatial tributary was sliced into 4.4-μs blocks. Simultaneously, TDM local oscillator (LO) light was generated by another SDM-TDM converter. The coherent Rx consisted of 70-GHz balanced photodiodes and a 70-GHz oscilloscope. The SUT was processed by offline digital signal processing (DSP). After chromatic dispersion compensation, it was demodulated using a frequency-domain 96×24 MIMO adaptive equalizer [17] with an FFT size of 2048. The demodulated signals were decoded assuming adaptive coding with forward error correction (FEC) to evaluate the net bitrates. The details of the decoding are the same as those shown in a previous work [18], except that FEC frames were constructed across all spatial channels to mitigate the effect of mode-dependent loss (MDL). The achievable bitrates were also evaluated based on generalized mutual information.

**Tab. 1:** Parameters of the transmission line.

| Parameter | Value |
| --- | --- |
| Fiber loss [dB/km] (excl. fan-in / fan-out) | 0.176 |
| Total span loss [dB] (incl. fan-in / fan-out) | 12.1 |
| Splicing / connector point number per span | 15 / 3 |
| Splicing loss per point [dB] | <0.1 |
| Connector loss per point [dB] | Avg. 0.12 dB |
| SMD in C-band [ps/km$^{1/2}$] | <6.0 |

**Results**

The experimental results are shown in Fig. 3. First, the transmission characteristics such as memory length (impulse response width) and MDL were investigated with WDM configuration. In this measurement, we use a channel with a center frequency of 194.1 THz. We defined the memory length $\tau_\mathrm{m}$ by the time window that encompassed 90 % of the power of the pulse energy, which are evaluated by the tap coefficient of the adaptive filter in the time domain. The results are shown in Fig. 3(a). $\tau_m$ was found to be limited to 2.1 ns even at a transmission distance of 1017 km. The results are well fitted by a curve of strongly coupled-core transmission $\tau_\mathrm{m} = a\sqrt{L}$, where $a = 67$ ps/km$^{1/2}$ is a fitting parameter and $L$ is the transmission distance. The discrepancy between $a$ and the fiber SMD coefficient could be attributed to the definition of $\tau_\mathrm{m}$ and remaining few cm level relative optical path length differences between the parallel recirculation loops. We also evaluated the rms MDL $\sigma_\mathrm{rms}$ [19] for each distance. By curve fitting, we estimated the rms MDL per span $\sigma_g$ as $\sigma_g = 0.35$ dB. We also performed 1-hour measurement for a distance of 428 km (8 spans) with 4-minute intervals. Both $\tau_\mathrm{m}$ and $\sigma_\mathrm{rms}$ were time-stable (average 1.34 ns / 1.98 dB, standard deviation 0.04 ns / 0.07 dB, respectively), even though tap coefficients, or, transmission matrices, from which $\tau_\mathrm{m}$ and $\sigma_\mathrm{rms}$ were estimated, varied with each measurement. This supports the feasibility of stable transmission in terrestrial field environments.

We then performed a full WDM measurement. The results are shown in Fig. 3(b), which shows the net bitrate and achievable bitrate for each WDM channel for both one span and 19 spans. In the 1-span measurements, all channels achieved more than 14 Tb/s/wavelength (average 14.69 Tb/s), while in the 19-span measurements, all channels exceeded 12 Tb/s/wavelength (average 12.55 Tb/s). The total net/achievable bitrate reached 455.4/465.8 Tb/s for one span and 389.3/409.0 Tb/s for 19 spans. $\tau_\mathrm{m}$ and $\sigma_\mathrm{rms}$ were also evaluated as shown in Fig. 3(c), showing a uniform characteristic across the entire C-band (average 1.70 ns / 2.52 dB, standard deviation 0.07 ns / 0.12 dB, respectively). The discrepancy of $\tau_m$ from the results shown in Fig. 3(a) might come from change of connection condition within the recirculation loops.

**Conclusion**

We presented the first demonstration of field-installed 12CCF transmission with >12 Tb/s spatial MIMO channels, demonstrating net 455.4 Tb/s at 53.5 km and 389.3 Tb/s at 1017 km with stable $\tau_\mathrm{m}$ and $\sigma_\mathrm{rms}$ in time and frequency. These results support the feasibility of a future SDM system deployment with sub-petabit-class total capacity and 10 Tb/s-class channel capacity.

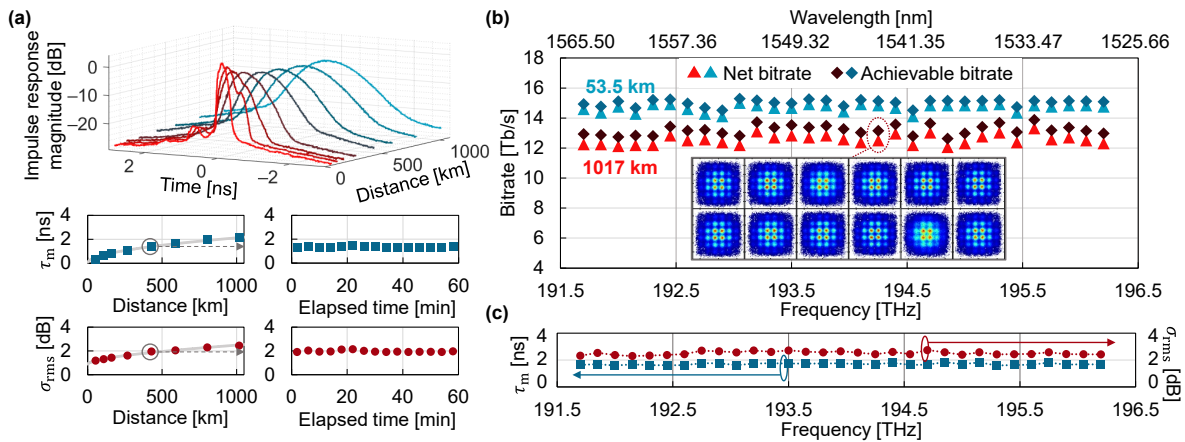

**Fig. 3:** Experimental results. (a) Impulse response magnitude versus distance. $\tau_m$ and $\sigma_\mathrm{rms}$ versus distance and elapsed time are also shown. (b) Bitrate obtained in WDM transmission. (c) $\tau_\mathrm{m}$ and $\sigma_\mathrm{rms}$ in WDM transmission.

## Acknowledgements

Part of this research is supported by National Institute of Information and Communications Technology (NICT) of Japan under the commissioned research JPJ012368C01001.